\documentclass[showpacs,10pt,twocolumn]{revtex4-1}

\usepackage{amsmath}
\usepackage{amssymb}
\usepackage{graphicx}
\usepackage{graphics}
\usepackage{epsfig}
\usepackage{CJK}
\usepackage{color}
\usepackage{epstopdf}
\usepackage{soul}

\usepackage{caption}

\begin{document}
\begin{CJK*}{GBK}{}

\title{Intriguing Kagome Topological Materials}
\author{Qi Wang$^{1,2,3}$, Hechang Lei$^{2,4,*}$, Yanpeng Qi$^{1,3,5,*}$, and Claudia Felser$^{6}$}
\affiliation{$^{1}$ School of Physical Science and Technology, ShanghaiTech University, Shanghai 201210, China\\
$^{2}$ Department of Physics and Beijing Key Laboratory of Opto-electronic Functional Materials $\&$ Micro-nano Devices, Renmin University of China, Beijing 100872, China\\
$^{3}$ ShanghaiTech Laboratory for Topological Physics, ShanghaiTech University, Shanghai 201210, China\\
$^{4}$ Key Laboratory of Quantum State Construction and Manipulation (Ministry of Education), Renmin University of China, Beijing 100872, China\\
$^{5}$ Shanghai Key Laboratory of High-Resolution Electron Microscopy, ShanghaiTech University, Shanghai 201210, China\\
$^{6}$ Max Planck Institute for Chemical Physics of Solids, Dresden 01187, Germany.\\
\textbf{Correspondence should be addressed to Y. P. Qi (qiyp@shanghaitech.edu.cn) or H. C. Lei (hlei@ruc.edu.cn)}
}

\date{\today}

\begin{abstract}
Topological quantum materials with kagome lattice have become the emerging frontier in the context of condensed matter physics. Kagome lattice harbors strong magnetic frustration and topological electronic states generated by the unique geometric configuration. Kagome lattice has the peculiar advantages in the aspects of magnetism, topology as well as strong correlation when the spin, charge, or orbit degrees of free is introduced, and providing a promising platform for investigating the entangled interactions among them.  
In this paper, we will systematically introduce the research progress on the kagome topological materials and give a perspective in the framework of the potential future development directions in this field.

\end{abstract}

\maketitle

\end{CJK*}

\textbf{Research progress on kagome materials}

The unique two-dimensional (2D) kagome lattice composed of corner-sharing triangles is a fascinating geometric configuration (Figure 1(a)). Decorating kagome lattice by adding atoms to its sites, which introduces spin, charge or orbital degrees of freedom, can engender abundant exotic quantum states. Initially, the studies primarily focused on the kagome insulating antiferromagnets.
It has been proposed that magnetic frustration exists in a 2D triangle lattice based on the Heisenberg antiferromagnet model. Naturally, the kagome lattice is considered to host a strong geometrical frustration effect, manifesting strong quantum fluctuations. It undoubtedly serves as one of the promising platform for studying the entangled quantum spin liquid (QSL) state under ultra-low temperatures. Cu-based kagome insulating antiferromagnet ZnCu$_{3}$(OH)$_{6}$Cl$_{2}$ was the first proposed QSL candidate \cite{13462,025003}. Various experimental and theoretical studies have been performed to elucidate the nature of QSL for a long period of time.

\begin{figure*}
	\centerline{\includegraphics[scale=0.66]{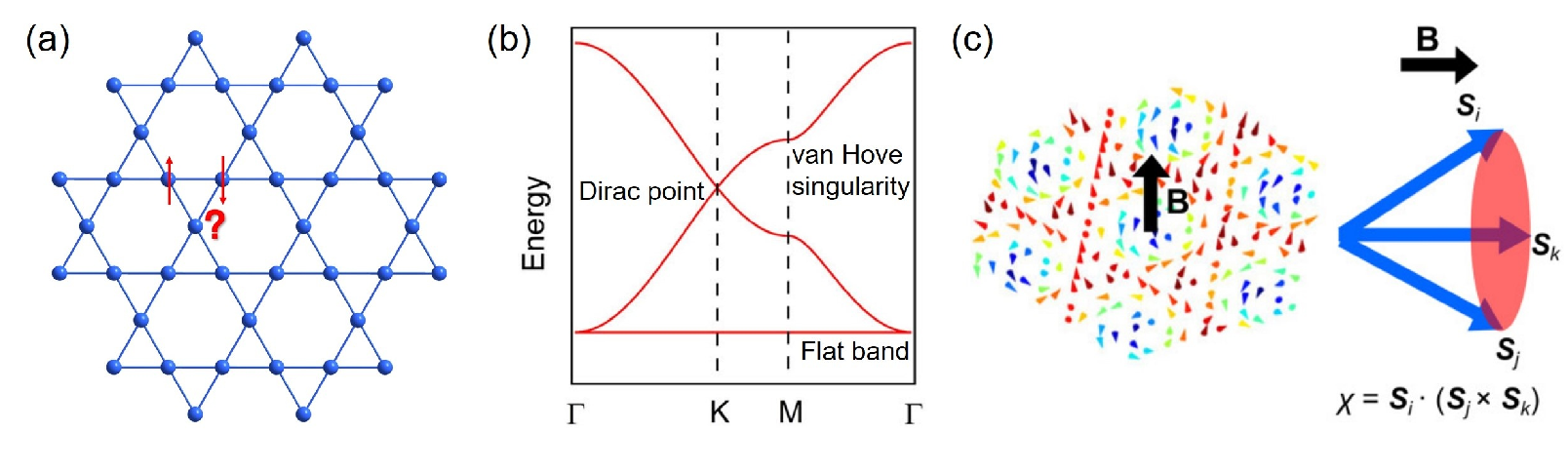}}
	\caption{ \textbf{The characteristics of kagome lattice.}
		(a) Geometric structure of a 2D kagome lattice. The red spins and question mark represent the schematic of geometric frustration on a triangle lattice. (b) Band structure of 2D kagome lattice in the absence of SOC in momentum space, featuring the Dirac point at the K point, van Hove singularities at the M point and a flat band. (c) Schematic of skyrmion lattice or noncoplanar spin texture with nonzero scalar spin chirality $\chi={\textit{\textbf{S}}}_{i}\cdot({\textit{\textbf{S}}}_{j}\times {\textit{\textbf{S}}}_{k})$ in real space. Adapted from Ref.\cite{014416,3129} }
\end{figure*}

Subsequently the studies on kagome insulators have been expanded to the field of topological kagome metals. It is theoretically predicted that the electronic band structure of 2D kagome lattice hosts nontrivial topological characteristics in the frame of the tight-bonding model without spin orbit coupling (SOC), featuring Dirac fermions with linear dispersion which are analogous to those in the honeycomb lattice, dispersionless flat bands as well as van Hove singularities (Figure 1(b)) \cite{786}.
On the one hand, the position of topological band crossings relative to the Fermi energy ($E_{\rm F}$) in kagome magnets with broken time-reversal symmetry plays a critical role in determining the magnitude of Berry curvature in momentum space. Especially when these crossings (Chern-gapped Dirac fermions or magnetic Weyl fermions) are close to the $E_{\rm F}$, the Berry curvature can be significantly enhanced, resulting in the generation of exotic electromagnetic responses of conduction electrons, such as a large intrinsic anomalous Hall effect (AHE) or even the quantum anomalous Hall effect (QAHE).
On the other hand, the van Hove singularities and flat bands in the kagome lattice could introduce significant electronic correlation effects due to the contribution of large density of states. It is possible to realize fascinating quantum states such as superconductivity, charge density wave (CDW), spin density wave (SDW), etc.
In addition, the frustrated kagome structure contributes to the generation of magnetic skyrmions or noncoplanar magnetic texture in real space (Figure 1(c)), offering opportunities to investigate the topological magnetic excitations, including the topological Hall effect (THE).

Experimentally, in 2016, Lei's group reported that the ferromagnetic Fe$_{3}$Sn$_{2}$ single crystal with Fe kagome lattice exhibits a large intrinsic AHE ($\sim$ 400 $\Omega^{-1}$cm$^{-1}$) \cite{075135}. Then in 2018, Checkelsky's group pointed out that the presence of massive Dirac fermions with a 30 meV gap near the $E_{\rm F}$, as revealed by angle-resolved photoemission spectroscopy (ARPES), is responsible for the large AHE due to the nonzero Berry curvature \cite{638}. 
In the same year, Lei's group and Felser's group independently identified the ferromagnetic Weyl semimetal Co$_{3}$Sn$_{2}$S$_{2}$ with Co kagome lattice experimentally \cite{3681,1125}. Thereinto, a giant intrinsic AHE ($\sim$ 500 - 1000 $\Omega^{-1}$cm$^{-1}$) and anomalous Hall angle associated with magnetic Weyl fermions slightly above $E_{\rm F}$, as well as negative magnetoresistance arising from chiral anomaly, were revealed. Subsequently, a series of spectroscopic experiments carried out on Co$_{3}$Sn$_{2}$S$_{2}$ by utilizing ARPES and high-resolution scanning tunneling microscopy (STM), further confirmed the existence of topological electronic states and unusual quantum phenomena \cite{1282,1278,443}. Furthermore, the antiferromagnets Mn$_{3}$X (X = Sn, Ge) with noncollinear spin configuration in the Mn kagome layer also exhibit a large AHE driven by the intrinsic Berry-phase mechanism \cite{212,e1501870}. For magnetic kagome compounds where the time-reversal symmetry is broken, the existence of such topological states as massive Dirac fermions or Weyl fermions in the vicinity of $E_{\rm F}$ in the condition of broken time-reversal symmetry and SOC effectively modulate the Berry curvature effect in momentum space, playing a vital role in the intriguing electromagnetic responses. Hence, the studies on the interplay between exotic magnetism and nontrivial band topology generate substantial interests in kagome magnets.

In the following years, a series of novel kagome magnets in which 3$d$ transition metal atoms form the kagome lattice were explored and investigated. 
In kagome antiferromagnet FeSn and paramagnet CoSn, experimental investigations confirmed the simultaneous existence of Dirac points and flat bands, which are rarely observed together in real kagome magnets \cite{163,4002,4003}. Another well-known class of kagome magnets is the RMn$_{6}$Sn$_{6}$ family (R = rare earth elements), which features a pristine Mn kagome lattice without the occupation of other atoms. This family exhibits various types of magnetic ground states by altering the R elements \cite{014416,246602,533,3129,644}. A quantum-limit Chern phase was realized in TbMn$_{6}$Sn$_{6}$ \cite{533}. The large AHE due to the Chern-gap-induced Berry curvature was observed in antiferromagnet YMn$_{6}$Sn$_{6}$ and ferrimagnet TbMn$_{6}$Sn$_{6}$ \cite{533,014416}. Moreover, YMn$_{6}$Sn$_{6}$ exhibits a large THE due to the nonzero scalar spin chirality that arises from the field-induced double-fan spin structure in real space \cite{014416}. It should be noted that the THE was also observed in Fe$_{3}$Sn$_{2}$ \cite{017101}. In contrast, the large THE possibly originates from the coexistence of skyrmionic bubbles and non-collinear spin textures. In addition to the nontrival Berry phase in momentum space, the contribution of real-space Berry phase arising from the magnetic frustrated kagome lattice also demonstrates significant impact on the transport properties.

Excitingly, since 2020, the discovery of V-based kagome superconductors AV$_{3}$Sb$_{5}$ (A = K, Rb, Cs) brings the study on the kagome lattice systems to a new climax \cite{094407,247002,037403,034801}. In contrast to the kagome magnets as mentioned above, this category of kagome metals, which lack long-range magnetic ordering, simultaneously accommodates the features of superconductivity (with a superconducting transition temperature $T_{\rm c}$ $\sim$ 0.92 - 2.5 K), CDW states and topological band structures \cite{247002,037403,034801,031026,031050,2102813,49}, providing a promising platform for studying the electronic correlation effects. In a short period of time, a substantial body of experimental and theoretical studies on the AV$_{3}$Sb$_{5}$ family have emerged, revealing exotic quantum phenomena such as the pressure-driven superconducting dome and reentrant superconductivity \cite{2102813,3645,247001}, non-trivial quantum oscillation \cite{207002}, a giant AHE \cite{eabb6003}, chiral CDW order \cite{1038}, pair density wave \cite{222}, nematic phase \cite{59} and so forth. Furthermore, the layered kagome structure with weak interlayer coupling, which can be easily cleaved, offers more possibilities for studying the low-dimensional physics within the kagome family \cite{105}. These investigations extremely enrich the physical properties of this system. 

The pace of the search for kagome superconductors has significantly accelerated following the discovery of AV$_{3}$Sb$_{5}$ family. For instance, inherent superconductivity observed in Ti-based kagome metal CsTi$_{3}$Bi$_{5}$ \cite{9626} and Ru-based kagome metals LaRu$_{3}$Si$_{2}$ and YRu$_{3}$Si$_{2}$ \cite{214527,087401}, as well as pressure-induced superconductivity in Pd-based kagome materials Pd$_{3}$P$_{2}$S$_{8}$, Pd$_{3}$Pb$_{2}$Se$_{2}$ and Rb$_{2}$Pd$_{3}$Se$_{4}$ \cite{043001,155115,123013,214501}.
Recently, the discovery of the unconventional superconductivity in Cr-based kagome antiferromagnet CsCr$_{3}$Sb$_{5}$ upon compression, which shares an identical structure with AV$_{3}$Sb$_{5}$, has further advanced the study of kagome superconductivity into a new phase \cite{1032}.

It has been theoretically proposed that the filling of van Hove in kagome lattice within considering the onsite Hubbard interaction and Coulomb interaction, could generate the exotic correlated electronic states such as CDW \cite{144402,115135,126405}. After revealing the competing CDW orders with superconductivity in AV$_{3}$Sb$_{5}$, which is closely associated with the van Hove singularities around $E_{\rm F}$, the CDW states have intensively garnered attention. The Fermi surface instability, loop current order, lattice degree of freedom are reported to contribute the CDW \cite{046401,036402,1384,3461}. It is intriguing that this provides a promising avenue for studying the electronic instability and various interactions with other correlated states in the topological kagome family. Furthermore, a nonmagnetic V-based kagome intermetallic ScV$_{6}$Sn$_{6}$ and Fe-based kagome antiferromagnet FeGe, were discovered to host CDW states \cite{216402,15,490,086902}. Currently, a growing number of novel kagome materials are being gradually discovered, such as AV$_{6}$Sb$_{6}$ (A = K, Rb, Cs), CsV$_{8}$Sb$_{12}$, Fe$_{3}$Ge \cite{127401,9823} and so on. 

\textbf{Perspective and conclusion}

The reported materials with kagome lattice demonstrate an extremely rich variety of quantum states of matter. Nevertheless, there still remain intriguing and significant quantum phenomena yet to be discovered, including but not limited to the following three aspects. 

(1) Metallization of QSL state

The QSL state, regarded as one of the important frontier fields in condensed matter physics, has attracted extensive attentions. However, only a few QSL candidate materials have been proposed and investigated experimentally. In contrast, the experimental realization of QSL insulators in kagome lattice systems is even more infrequent. Currently, the existence of the QSL state under ultra-low temperatures still remains controversial due to the lack of direct experimental evidence. More novel kagome QSL candidates are required to identify the QSL state. 

On account of the characteristics of the insulating ground state in QSL, it offers the potential to modulate the electronic structure or chemical potential to tune the insulating behavior and make it metallized. Particularly, it is proposed that the QSL state can be manipulated to access high-temperature superconductivity by doping with charge carriers \cite{1196}. 
In addition to chemical doping, high-pressure regulation utilizing the diamond anvil cell (DAC) technique is also a widely employed tuning method in terms of its clean feature, which do not introduce impurities or defects. Currently, a few QSL candidates have been found to exhibit an insulator-metal transition under pressure regulation. The pressure-driven superconductivity is relatively rare and has only been observed in NaYbSe$_{2}$ with triangle lattice \cite{097404}. However, there have been no successful realization of the metallization or superconductivity in kagome QSL to date, such as the chemical doping in ZnLi$_{x}$Cu$_{3}$(OH)$_{6}$Cl$_{2}$ and Ga$_{x}$Cu$_{4-x}$(OH)$_{6}$Cl$_{2}$ \cite{041007,1800663}. On the basis of the enrichment of the novel kagome QSL candidates, it is expected to achieve the metallization or superconductivity within the quantum regulation in the future. Moreover, the realization of superconducting state will promote the insights into the mechanisms of high-temperature superconductivity.

(2) Topological transport quantization

It is generally accepted that various particular quantum states will emerge in the low-dimentional physics, such as the QAHE.
The QAHE, characterized by a quantized anomalous Hall conductance, represents the quantum Hall effect in the absence of magnetic field \cite{2015}.
It was first experimentally observed in Cr-doped magnetic topological insulator (Bi,Sb)$_{2}$Te$_{3}$ in the 2D limit.\cite{167} The exploration of high-temperature QAHE in  the intrinsic magnetic topological insulators has consistently been the focus of the research.
2D magnetic kagome lattice system has been predicted to be a promising platform to realize the QAHE \cite{R6065,365801,186802}.
In the presence of magnetic ordering and SOC in 2D limit, a topologically nontrivial gap opens at the band crossings in momentum space. In particularly, the QAHE emerges with nontrivial Chern number when the $E_{\rm F}$ locates within the insulating gap. For example, Mn-based ferromagnetic insulator Cs$_{2}$LiMn$_{3}$F$_{12}$ with Chern number = 1 in the condition of the single layer and thin film \cite{186802}, Co-based ferromagnets Co$_{3}$A$_{2}$B$_{2}$ (A = Sn, Pb; B = S, Se) in the 2D limit with high Chern number = 3 or 6 \cite{115106,014410}. Experimentally, although the large intrinsic AHE has been intensively studied in various kagome magnets, the magnitude of the anomalous Hall conductivity remains smaller than the order of $e^{2}/h$, i.e., the QAHE has not yet been achieved in real magnetic kagome materials. 
	
The achievement of magnetic kagome layer in 2D limit is extremely difficult mainly due to the strong interlayer couplings or intralayer interactions in the currently reported kagome magnets. It is difficult to cleave this category materials into low-dimentional ones. The realization of magnetic layered materials with perfect 2D magnetic kagome lattice is excepted to achieve this peculiar low-dimentional quantum phenomena.

(3) Unconventional Superconductivity

In recent years, the researches on the magnetism and superconductivity of kagome materials have made significant progress. Nevertheless, it is uncommon to realize a superconducting state in magnetic kagome materials, which is expected to reveal quantum criticality and unconventional superconductivity. Currently, the interplay between magnetism and superconductivity, the mechanism underlying unconventional superconducting pairing, as well as the impact of spin fluctuations on superconductivity remains inadequately understood and requires further investigation. Furthermore, in view of nontrivial topological band structure in kagome lattice, it provides a perfect opportunity to study the topological superconductivity characterized by a nontrivial topological invariant.
Exploring the inherent kagome topological superconductors or realizing the superconducting states in topological kagome materials by means of physical/chemical pressure, gating, etc. It provides more opportunities to achieve Majorana zero modes and topological quantum computation. More novel kagome superconductors and nontrivial quantum phenomena remain to be excavated.

In conclusion, the kagome lattice systems have triggered significant attentions due to the emergent exotic quantum states, making it the focus of research within condensed matter physics. The diversity of kagome families and tunability of various degrees of freedom also provide multiple feasibilities for systematic research. In the future, the achievements of the scenario in the aspect of topological quantization and unconventional superconductivity in kagome systems possibly will provide promising applications.

\textbf{ACKNOWLEDGEMENTS}

This work was supported by the National Natural Science Foundation of China (Grant Nos. 12404161, 52272265, 12274459), the National Key R\&D Program of China (Grant Nos. 2023YFA1607400, 2023YFA1406500, 2022YFA1403800), and the Shanghai Sailing Program (Grant No. 23YF1426800).

\textbf{AUTHOR CONTRIBUTIONS}

Q.W., H.L., Y.Q., and C.F. wrote the manuscript.

\textbf{COMPETING INTERESTS}

The authors declare no competing interests.

\end{document}